# Observation of Single Top Quark Production at D0

Liang Li

*Department of Physics and Astronomy, University of California, Riverside, CA 92521, USA*

**Abstract.** This paper presents the observation of the electroweak production of single top quarks in the D0 detector at the Fermilab Tevatron $p\bar{p}$ Collider at a center-of-mass energy of 1.96 TeV. Events containing an isolated electron or muon and missing transverse energy, together with jets originating from the fragmentation of b quarks are used to measure a cross section for single top-quark production of $\sigma(p\bar{p} \to tb + X, tqb + X) = 3.94 \pm 0.88$ pb. The probability to measure a cross section at this value or higher in the absence of signal is $2.5 \times 10^{-7}$, corresponding to a 5.0 standard deviation significance.

## INTRODUCTION

The top quark, discovered at Tevatron collider at Fermi National Accelerator Laboratory in 1995 [1], is by far the heaviest elementary particle found to date. The standard model (SM) predicts two independent top quark production mechanisms at hadron colliders: the primary process in which top quarks are produced in pair via the strong interaction and the second process in which the top quark is produced singly via the electroweak interaction. The D0 and CDF collaborations observed top quarks via pair production, however the single top production process has not been observed until the completion of this analysis [2]. The two main modes of single top production [3] are: the *s*-channel process (via the decay of a virtual *W* boson, referred as "*tb*"), and the *t*-channel process (via the exchange of a virtual *W* boson between a light quark and a bottom quark, referred as "*tqb*"). The sum of their predicted cross sections is $3.46 \pm 0.18$ pb [4] for a top quark mass of 170 GeV. Studying the electroweak production allows one to directly probe the *Wtb* vertex and measure the magnitude of the Cabibbo-Kobayashi-Maskawa (CKM) matrix element $|V_{tb}|$ without assuming three quark generations or CKM unitarity. The same analysis method can also be used to identify other rare physics processes, such as the Higgs boson production.

## EVENT SELECTION AND MODELING

The events used in this analysis are selected from 2.3 fb$^{-1}$ of data recorded using the D0 detector [5] between 2002 and 2007. Events are selected containing exactly one isolated high $p_T$ electron or muon, missing transverse energy, and at least two jets, with at least one jet being identified as originating from the fragmentation of a $b$ quark ($b$-jet) and requiring the $\not{E}_T$ is not along the direction of the lepton or the leading jet and a minimum total transverse energy $H_T$ to suppress the QCD multijet background.

Single top signal events are modeled using the COMPHEP-based next-to-leading order (NLO) Monte Carlo (MC) event generator SINGLETOP [6] assuming SM production for the ratio of the $tb$ and $tqb$ cross sections. The dominant $W$+jets background, the $Z$+jets and $t\bar{t}$ backgrounds are simulated using the ALPGEN leading-log MC event generator [7] using PYTHIA [8] to model hadronization. The small diboson backgrounds are modelled using PYTHIA. All MC events are passed through a GEANT-based simulation of the D0 detector. Small differences between data and simulation in the lepton and jet reconstruction efficiencies and resolutions are corrected in the simulation. Additional correction for the η(jets), Δη (jet1, jet2), and Δϕ (jet1, jet2) distribution in the $W$+jets samples are applied to match data. The multijets background is modeled using independent data samples containing leptons that are not isolated. The fraction of $W/Z$+jets events containing heavy flavor jets is normalized to the NLO value [9]. The numbers of expected $W$+jets and QCD multijet events are normalized to the data sample with other backgrounds subtracted before $b$-tagging. The $t\bar{t}$ background is normalized to the predicted cross section [10]. After $b$-tagging, an empirical correction factor of 0.95±0.13 derived from the $b$-tagged and not-$b$-tagged two-jet data and simulated samples is applied to the $Wbb$ and $Wcc$ samples to correct for higher-order effects. To improve the sensitivity, the selected events are then divided into 24 separate analysis channels. The above selections give 4,519 events, which are expected to contain $223\pm30$ single top quark events. Table 1 shows the event yields, separated by jet multiplicity. Good agreement is found between data and predictions in the signal sample and the two cross-check samples used to check the modeling of the two main backgrounds: $W$+jets and $t\bar{t}$. Figure 1 shows the $W$ transverse mass distribution for all 24 channels combined.

**TABLE .** Number of expected and observed events in 2.3 fb$^{-1}$ separated by jet multiplicity.

| Source | 2 jets | 3 jets | 4 jets |
|---|---|---|---|
| $tb+tqb$ signal | 139 ± 18 | 63±10 | 21±5 |
| $W$+jets | 1,829 ± 161 | 637±61 | 180±18 |
| $Z$+jets and dibosons | 229 ± 38 | 85±17 | 26±7 |
| $t\bar{t}$ | 222 ± 35 | 436±66 | 484±71 |
| Multijets | 196 ± 50 | 73±17 | 30±6 |
| Total prediction | 2,615 ± 192 | 1,294±107 | 742±80 |
| Data | 2,579 | 1,216 | 724 |

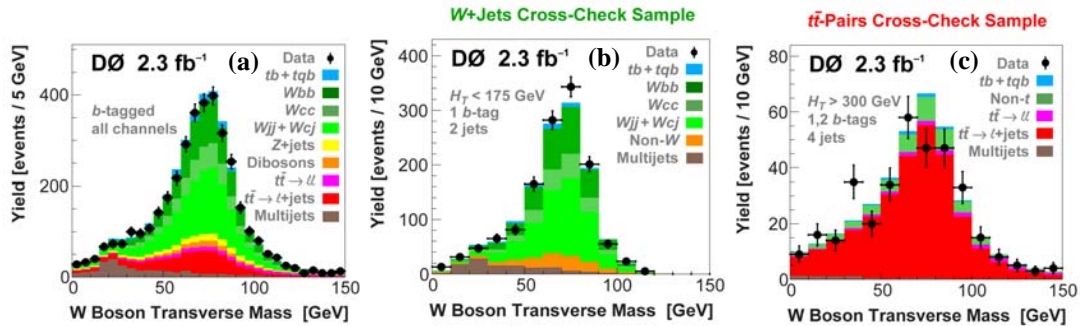

**FIGURE 1.** *W* Boson transverse mass distribution after all selection (all channels combined) for the signal sample (a) and for the *W*+jets (b) and $t\bar{t}$ (c) cross-check samples.

Systematic uncertainties are considered for all corrections applied to the background model. The largest uncertainties come from the jet energy scale and the tag-rate functions, with smaller contributions from MC statistics, the correction for jet-flavor composition in *W*+jets events, and from the background normalizations. The total uncertainty on the background is 8%–16% depending on the analysis channel.

## MULTIVARIATE ANALYSIS

Three independent multivariate analysis (MVA) techniques are used to separate the small single top signal from the large backgrounds: boosted decision trees (BDT) [11], Bayesian neural networks (BNN) [12], and the matrix element (ME) method [13]. The implementation of these MVA techniques is improved from the previous analysis [14, 15] in the choice of input variables and some detailed tuning of each technique.

The three multivariate techniques use the same data sample but are not completely correlated. A combination BNN (BNNComb) trained by using the three individual discriminant outputs as inputs increases the expected sensitivity and the precision of the cross section measurement. MVA techniques are tested by generating ensembles of pseudodatasets created from background and signal at different cross sections to confirm a linear response and an unbiased cross section measurement. The output discriminants have also been produced for the cross-check samples and demonstrate that the backgrounds are well-modeled across the full range of the discriminant output.

## RESULTS

The single top production cross section is measured using the same Bayesian approach as in the previous analysis [14, 15]. This is a Bayesian calculation forming a binned likelihood as a product over all bins and channels using a flat non-negative prior for the cross section. The position of the peak of the resulting posterior density gives the cross section value, and the 68% interval about the peak gives the uncertainty. The sensitivity of each analysis to a contribution from single top quark production is estimated by generating ensembles of pseudodatasets that sample the background model and its uncertainties in the absence of signal. A cross section is measured from each pseudodataset, which allows us to calculate the probability to measure the SM cross section ("expected significance") or the observed cross section ("observed significance"). Table 2 summarizes the cross section measurement results. The cross section measured by the BNNComb has a *p*-value of $2.5 \times 10^{-7}$ and a significance of 5.0 standard deviation, thus the unambiguous observation of single top production. Figure 3 shows the distribution of the combination output and examples of variables with high sensitivity to the signal, which illustrates the importance of the signal to achieve a good modeling of the data. The cross section measurement is then used to determine the Bayesian posterior for $|V_{tb}|^2$ in the interval [0,1] and extract a limit of $|V_{tb}| > 0.78$ at 95% C.L. within the SM without assuming three generations of quarks [6]. When the upper constraint is removed, the measurement gives $|V_{tb}| = 1.07 \pm 0.12$, where $f_1^L$ is the strength of the left-handed $Wtb$ coupling.

**TABLE.** Cross section measurements, expected and observed significances for MVA techniques.

| MVA | $\sigma \pm \Delta\sigma$ (pb) (exp.) | $\sigma \pm \Delta\sigma$ (pb) (obs.) | Expected (SD) | Observed (SD) |
|---|---|---|---|---|
| BDT | $3.61^{+0.95}_{-0.89}$ | $3.74^{+0.95}_{-0.79}$ | 4.3 | 4.6 |
| BNN | $3.60^{+1.02}_{-0.90}$ | $4.70^{+1.18}_{-0.93}$ | 4.1 | 5.4 |
| ME | $3.60^{+1.10}_{-0.96}$ | $4.30^{+0.99}_{-1.20}$ | 4.1 | 4.9 |
| BNNComb | $3.50^{+0.99}_{-0.77}$ | $3.94 \pm 0.88$ | 4.5 | 5.0 |

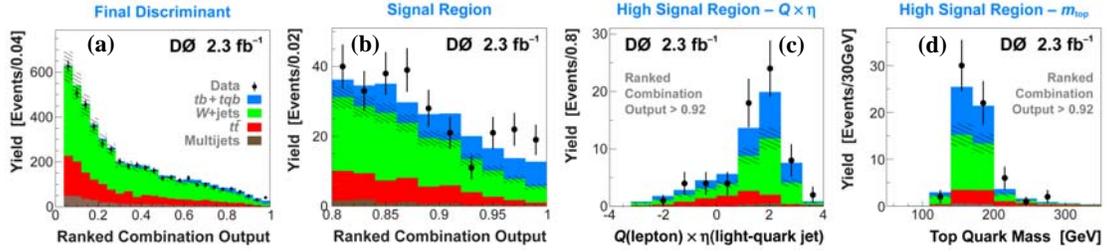

**FIGURE 2.** Distribution of the combination output for all 24 analysis channels combined, (a) full range, and (b) high signal region. The bins have been ordered by their expected signal-to-background ratio and the signal is normalized to the measured cross section. The hatched band indicates the total uncertainty on the background. Distribution of lepton charge times pseudorapidity of the leading not-b-tagged jet (c), and the reconstructed top quark mass (d) for ranked combination discriminant > 0.92.